\documentclass{svproc}

\pdfoutput = 1

\usepackage{url}

\usepackage{graphicx}
\usepackage{subfig}
\usepackage[usenames, dvipsnames]{color}
\usepackage[noadjust]{cite}
\usepackage{paralist}
\usepackage{hyperref}

\begin{document}
\mainmatter              
\title{Relative Importance of Convective Uncertainties}
\titlerunning{}  
%
\author{Etienne A. Kaiser\inst{1}\href{https://orcid.org/0000-0001-7237-2960}{\includegraphics[scale=0.6]{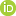}} \and Raphael Hirschi\inst{1,2}\href{https://orcid.org/0000-0001-8764-6522}{\includegraphics[scale=0.6]{figures/orcid_16x16.png}} \and W. David Arnett\inst{3}\href{https://orcid.org/0000-0002-6114-6973}{\includegraphics[scale=0.6]{figures/orcid_16x16.png}} \and Andrea Cristini\inst{4}\href{https://orcid.org/0000-0001-7299-4364}{\includegraphics[scale=0.6]{figures/orcid_16x16.png}} \and Cyril Georgy\inst{5}\href{https://orcid.org/0000-0003-2362-4089}{\includegraphics[scale=0.6]{figures/orcid_16x16.png}} \and Laura  J. A. Scott\inst{1}\href{https://orcid.org/0000-0001-8696-7981}{\includegraphics[scale=0.6]{figures/orcid_16x16.png}}}
\authorrunning{Kaiser et al.} 
%
\tocauthor{Etienne Kaiser, Raphael Hirschi, W. David Arnett, Andrea Cristini, Cyril Georgy and Laura Scott}
\institute{Keele University, Keele, UK\\
\email{e.kaiser@keele.ac.uk}
\and
WPI Kavli-IPMU, University of Tokyo, Japan
\and
University of Arizona, USA
\and
University of Oklahoma, USA
\and
University of Geneva, Geneva, Switzerland}

\maketitle              

\begin{abstract}
Convection plays a key role in the evolution of stars due to energy transport and mixing of composition. Despite its importance, this process is still not well understood. One longstanding conundrum in all 1D stellar evolution codes is the treatment of convective boundaries. In this study we compare two convective uncertainties, the boundary location (Ledoux versus Schwarzschild) and the amount of extra mixing, and their impact on the early evolution of massive stars. With increasing convective boundary mixing (CBM), we find a convergence of the two different boundary locations, a decreasing blue to red super giant ratio and a reduced importance of semiconvection.
\end{abstract}
\section{Introduction and Methodology}
In the framework of the mixing-length theory (MLT) \cite{Boehm-Vitense1958}, the location of the convective boundary is not defined and has to be determined by either the Ledoux or the Schwarzschild criterion. In regions with a gradient in chemical composition, the two criteria differ, leading to a region mixed by semiconvection. The process responsible and the efficiency of this mixing is a matter of debate.\\
Another source of significant uncertainty emerges from the treatment of CBM which is not included in MLT. Several add-ons have been proposed to account for the mixing of the boundary region, however, CBM is still an open question.\\
\\
We calculated $15\,M_\odot$, non-rotating stellar models at solar metallicity using the MESA stellar evolution code \cite{Paxton2011, Paxton2013, Paxton2015, Paxton2018}. All stellar models are computed several times, once with the Schwarzschild and three times with the Ledoux criterion, the latter with varied semiconvective efficiency. In the Ledoux models, we use the semiconvective formalism from \cite{Langer1983} (their Eq.\,(10)), with the adjustable efficiency parameter $\alpha_{\rm{sc}}$. In all models, we chose the exponentially decaying diffusive overshoot description (Eq.\,(2) from \cite{Herwig2000}) to account for CBM, with the adjustable parameter $f_{\rm{CBM}}$ which determines the length scale of CBM in pressure scale height.

\section{Discussion and Conclusions}
\begin{figure}[t!]
\subfloat[Ledoux, $f_{\rm{CBM}}=0.004$, $\alpha_{\rm{sc}}=0.04$]{\includegraphics[width=0.5\linewidth]{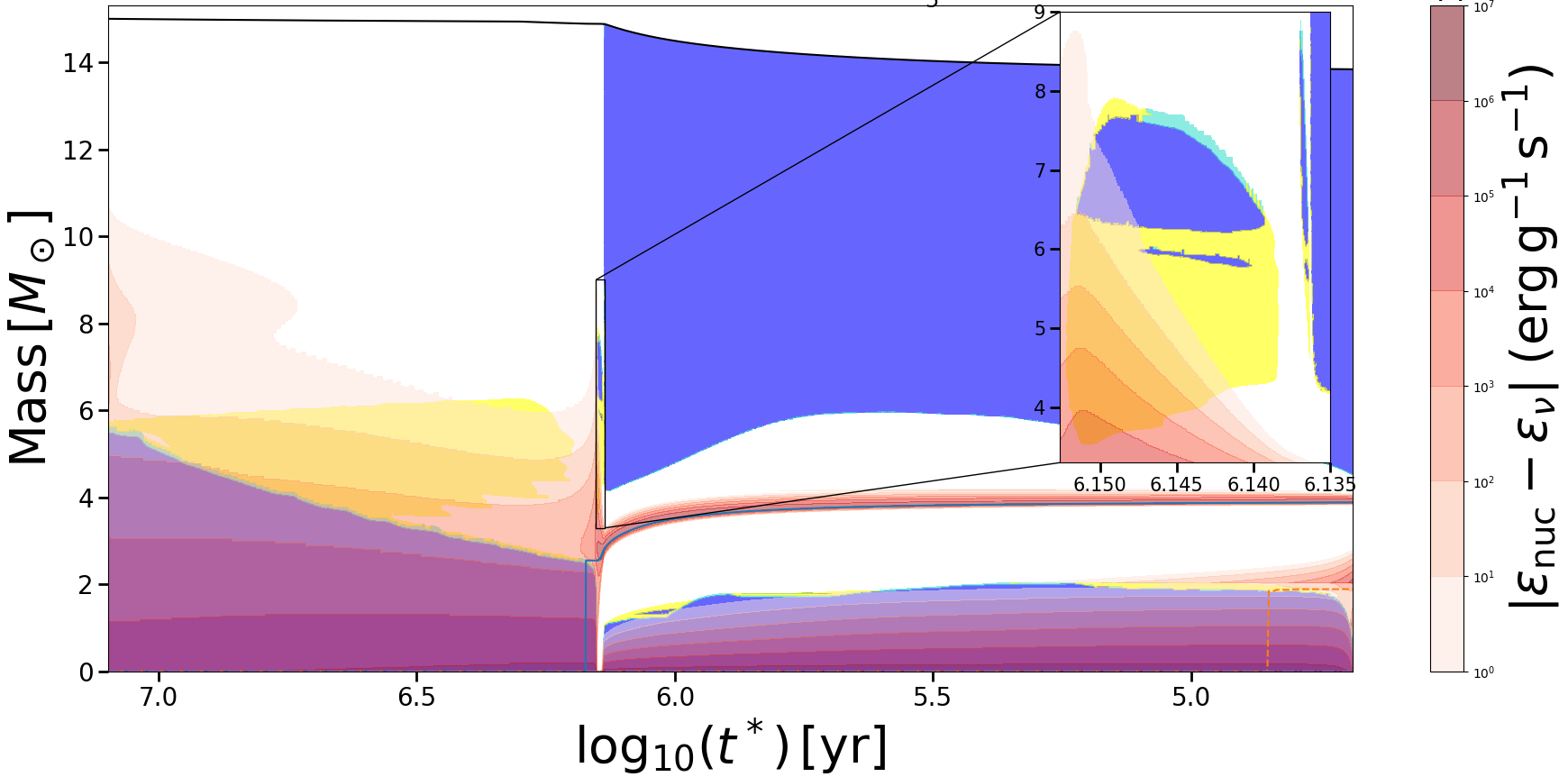}}
\subfloat[Schwarzschild, $f_{\rm{CBM}}=0.004$]{\includegraphics[width=0.5\linewidth]{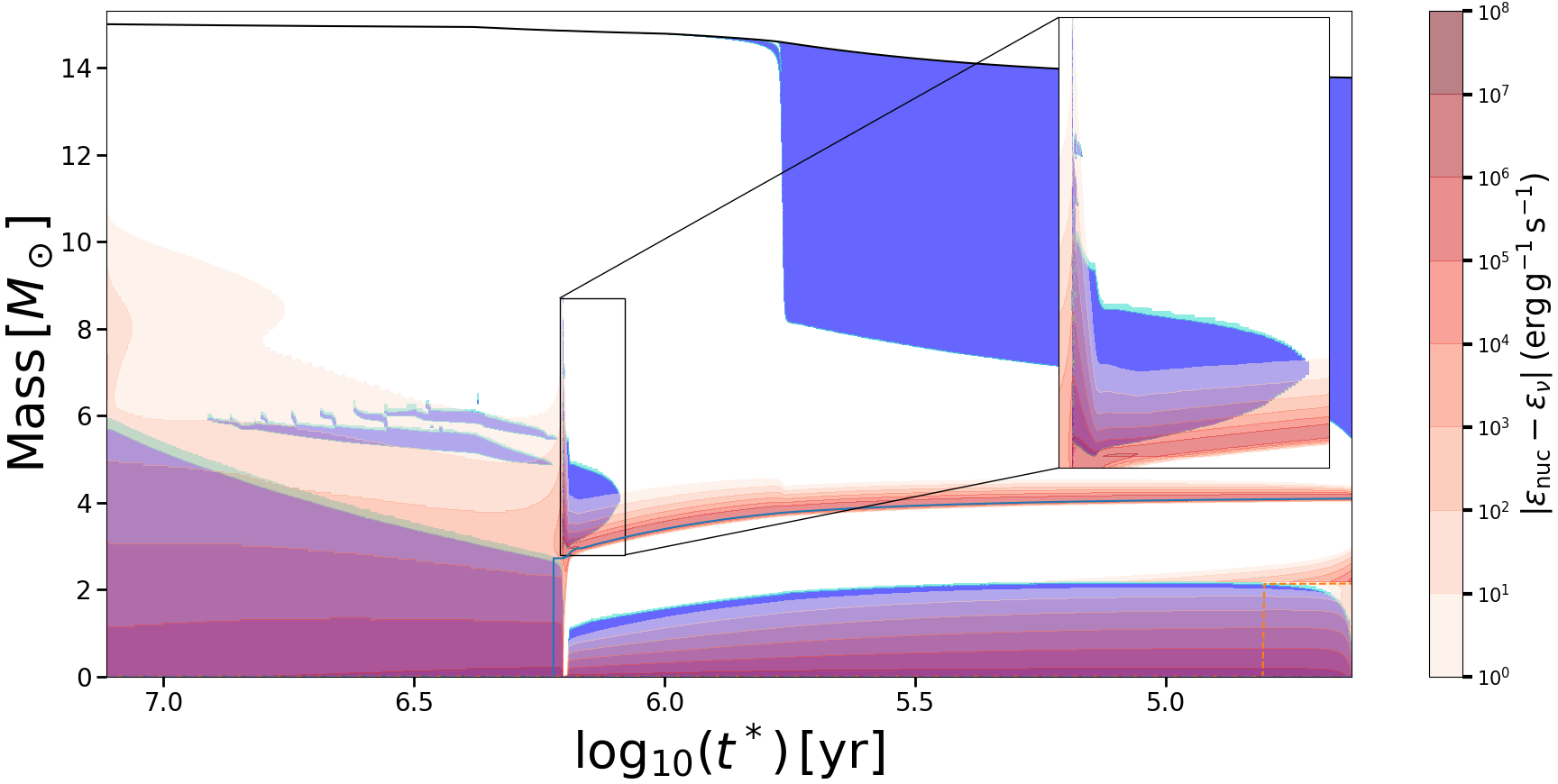}}
\qquad
\subfloat[Ledoux, $f_{\rm{CBM}}=0.022$, $\alpha_{\rm{sc}}=0.04$]{\includegraphics[width=0.5\linewidth]{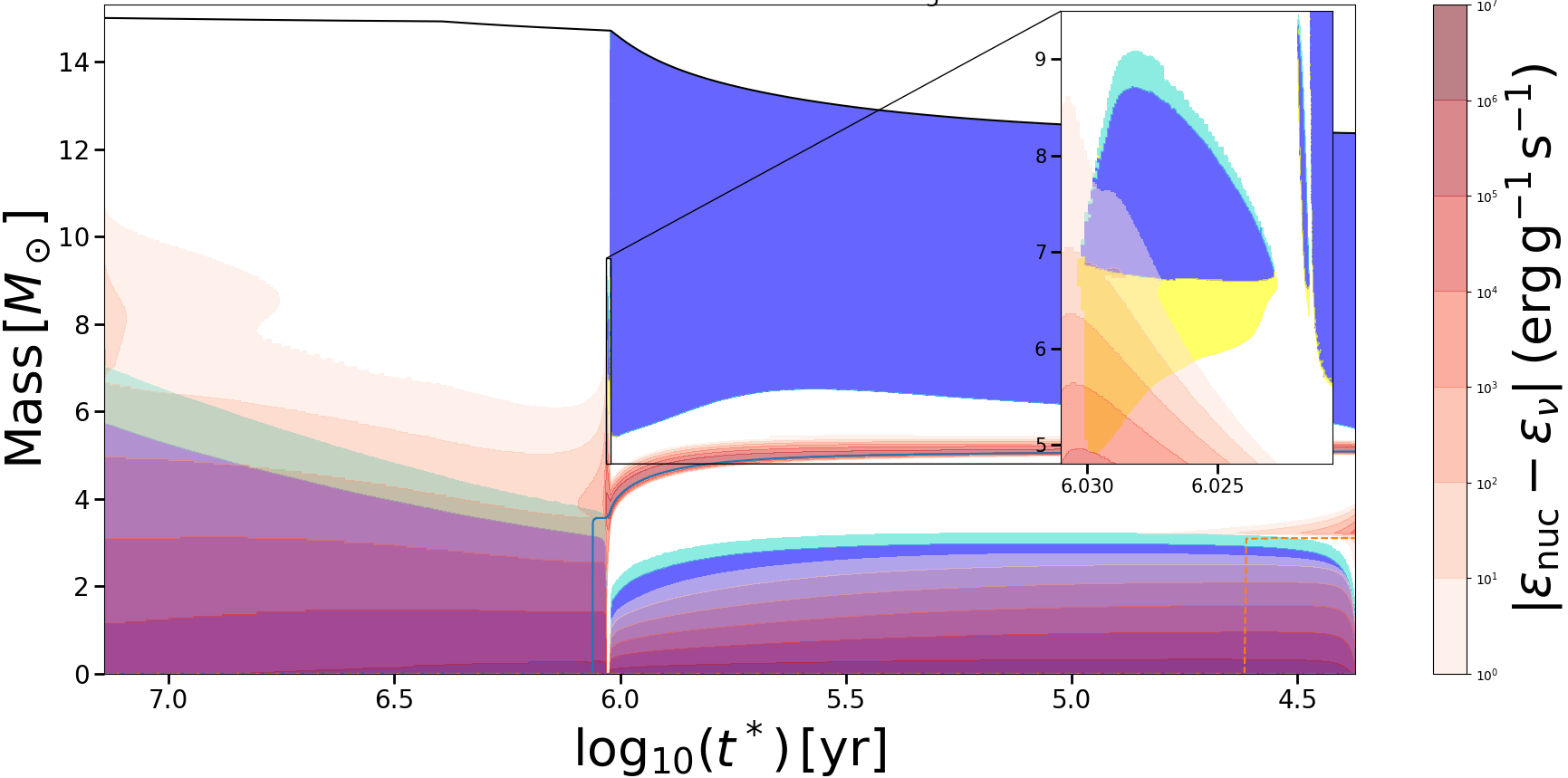}}
\subfloat[Schwarzschild, $f_{\rm{CBM}}=0.022$]{\includegraphics[width=0.5\linewidth]{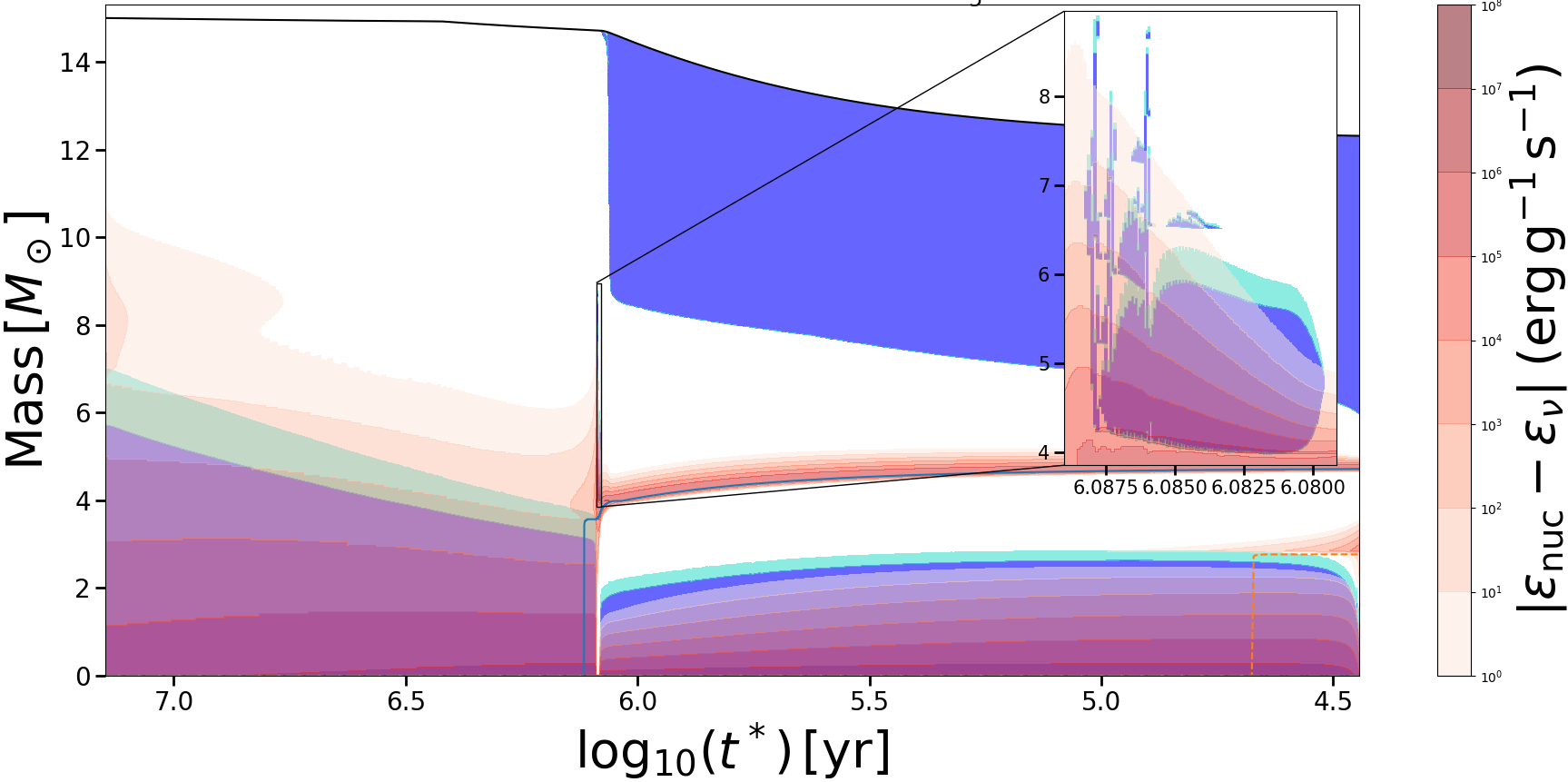}}
\caption{Structure evolution diagrams, illustrating the convective and burning history until helium depletion of some of the stellar models. The x-axis shows the time left until the star begins to collapse. Blue shading indicates convective zones, turquoise shading the CBM region, yellow shading the semiconvective zones and the red gradient the nuclear energy generation rate (minus neutrino losses). The zoom shows the intermediate convective zone which is crucial to determine whether the model stays in the blue super giant (BSG) or crosses directly to the red super giant (RSG) branch.}\label{Kippenhahn}
\end{figure}

In general, the extra mixing, either by CBM or semiconvection, smooths out the gradients in temperature and chemical composition, stabilizing the region above the hydrogen core against dynamical and vibrational instabilities, as shown in Fig.\,\ref{Kippenhahn}.
This reduces the differences between the two boundary criteria and they start to converge. Further distinctions are:
\begin{compactenum}[i]
\item CBM starts to appreciably change the structure already for a small value of $f_{\rm{CBM}}$ ($>0.004$).
\item The greatest influence of semiconvection is for no or very weak CBM. 3D hydrodynamic simulation indicate a non-negligible amount of CBM \cite{Cristini2017}, reducing the occurrence and thus the importance of semiconvection.
\item The size of the helium core, which is an indicator of the further stellar evolution, depends on the strength of the extra mixing. Generally, a higher amount of extra mixing results in larger helium cores.
\item The extra mixing increases the main-sequence width. Moreover, all the Ledoux models, and the Schwarzschild models with large amounts of CBM, cross the Herzsprung-Russel diagram directly towards the RSG branch whereas Schwarzschild models with no or less CBM spend time in the BSG branch.
\end{compactenum}
This study illustrates the need to determine the amount of CBM, which might reduce the discrepancy in stellar evolution due to the different boundary locations. Moreover, the time a star spends in the BSG branch before entering the RSG branch depends strongly on the amount of extra mixing and the stability criterion. This has crucial influence on the mass loss. These uncertainties can be tackled with 3D hydrodynamic simulations and asteroseismology (e.g. \cite{Cristini2017, Arnett2015, Arnett2017}). We will investigate the impact on nucleosynthesis and the advanced burning stages in a future work.

\paragraph{Acknowledgement}
The authors acknowledge support from the ChETEC COST Action (CA16117), supported by COST (European Cooperation in Science and Technology). This research has made use of the NASA's Astrophysics Data System Bibliographic Services. RH and CG thank ISSI Bern for meeting support.
%
%

\end{document}